\documentclass{gistt}
\usepackage{graphicx} % Required for inserting images
\usepackage{paralist}

\usepackage{listings}
\usepackage{tikz}
\usetikzlibrary{shapes,arrows}
\usetikzlibrary{calc}
\usetikzlibrary{fit,positioning}
\usetikzlibrary{patterns}

\title{Interoperability From Kieker to OpenTelemetry:\\ Demonstrated as Export to ExplorViz}
\author{David Georg Reichelt\\
Lancaster University Leipzig \& \\
Universität Leipzig
\and
Malte Hansen\\
Kiel University\\
\and
Shinhyung Yang\\
Kiel University\\
\and
Wilhelm Hasselbring\\
Kiel University
}

\bibliography{quellen.bib}

\begin{document}

\maketitle

\begin{abstract}
While the observability framework Kieker has a low overhead for tracing, its results currently cannot be used in most analysis tools due to lack of interoperability of the data formats. The OpenTelemetry standard aims for standardizing observability data. 

In this work, we describe how to export Kieker distributed tracing data to OpenTelemetry. This is done using the pipe-and-filter framework TeeTime. For TeeTime, a stage was defined that uses Kieker execution data, which can be created from most record types. We demonstrate the usability of our approach by visualizing trace data of TeaStore in the ExplorViz visualization tool.

\end{abstract}

\section{Introduction}

Observability data are used to understand a system's behavior. OpenTelemetry has become the de-facto standard for observability data~\cite{Blanco2023}. It offers a variety of integrations into different languages, libraries and frameworks for obtaining observability data. Storage and analysis tools like Zipkin, Prometheus, Grafana, and ExplorViz~\cite{hasselbring2020explorviz} rely on usage of the OpenTelemetry format, and enable understanding the system behavior by automated analysis and visualizations. While the observability framework Kieker~\cite{hasselbring2020kieker} has lower performance overhead for gathering distributed execution traces~\cite{reichelt2024overhead, reichelt2021overhead}, its data format is not usable for standard observability analysis tools. Therefore, creating interoperability between Kieker and OpenTelemetry will allow the usage for Kieker with its low overhead while maintaining Kieker's advantages. 

In this paper, we
\begin{inparaenum}[(1)]
  \item describe our concept for interoperability between Kieker and OpenTelemetry,
  \item describe our first step towards interoperability, which is the implementation of the export from Kieker to the OpenTelemetry format, and
  \item demonstrate the viability of this concept by visualizing the Kieker-monitored runtime behavior of the TeaStore~\cite{Eismann2018} in ExplorViz, which is achieved using the Kieker to OpenTelemetry export.
\end{inparaenum}

The remainder of this paper is structured as follows: First, we describe a concept for interoperability between Kieker and OpenTelemetry. Afterwards, we describe how the export of Kieker traces into OpenTelemetry traces can be accomplished. Subsequently, we describe how the export from Kieker to OpenTelemetry is implemented. This is demonstrated by using Kieker data for an ExplorViz visualization. Afterwards, we compare this approach to related work. Finally, we give a summary of our work.

\section{Concept for Interoperability Between Kieker and OpenTelemetry}
Kieker includes two parts: monitoring and analysis. The Kieker monitoring
part generates various observability data, including traces in its own record types. The Kieker analysis part
consumes Kieker traces for different analytics. 
OpenTelemetry is an ongoing effort to provide language- and
vendor-agnostic observability that became popular among Cloud Native Computing Foundation projects.\footnote{\url{https://landscape.cncf.io/}} However,
Kieker record types and OpenTelemetry formats are not one-to-one compatible;
thus, making the Kieker analysis part incapable of analyzing
OpenTelemetry-instrumented applications (and vice versa). %It is the same for the analytics that
%only ingest OpenTelemetry data; they cannot read Kieker records.

SustainKieker,\footnote{Funded by the Deutsche Forschungsgemeinschaft (DFG -- German Research Foundation), grant no. 528713834.} has specified two goals to address the
interoperability between Kieker and OpenTelemetry: the interoperability
\begin{inparaenum}[(1)]
\item should be bidirectional, and
\item should be transparent to the users.
\end{inparaenum}
Existing Kieker and OpenTelemetry analytics should be able to use Kieker and
OpenTelemetry data seamlessly without further effort. Our first effort to
export Kieker records in OpenTelemetry format for ExplorViz motivates our
approach to interoperability between the two systems. We describe the 
details in the following.

\section{Export Kieker to OpenTelemetry}

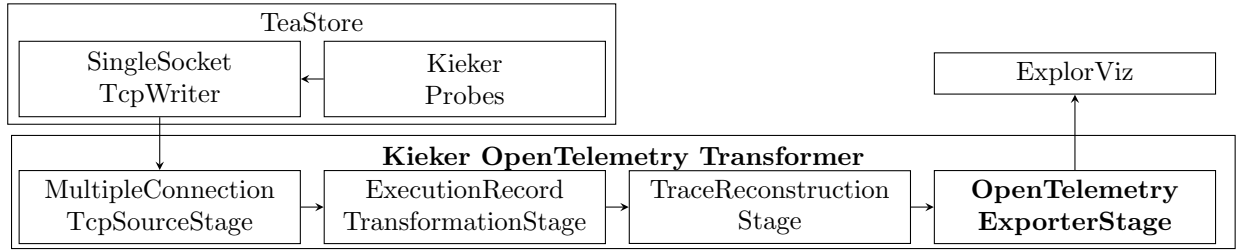
\begin{figure*}[h]
  \hspace{-0.5cm}
  \begin{center}
  \begin{tikzpicture}[>=stealth, node distance=0.3cm, every node/.style={rectangle, draw, align=center, minimum width=3.7cm, minimum height=0.5cm}]

  \node (TeaStore) [minimum width=8cm] {TeaStore\\[0.7cm]};
  \node (SingleSocketTcpWriter)[above=0.1cm of TeaStore.south, xshift=-2cm, minimum height=1.0cm] {SingleSocket\\TcpWriter};
  \node (Probes)[right=of SingleSocketTcpWriter, minimum height=1.0cm] {Kieker\\ Probes};
  
  \node (MultipleConnectionTcpSource) [below=0.7cm of SingleSocketTcpWriter] {MultipleConnection\\TcpSourceStage};
  \node (ExecutionRecordTransformation) [right=of MultipleConnectionTcpSource] {ExecutionRecord\\TransformationStage};
  \node (TraceReconstruction) [right=of ExecutionRecordTransformation] {TraceReconstruction\\Stage};
  \node (OpenTelemetryExporter) [right=of TraceReconstruction] {\textbf{OpenTelemetry}\\\textbf{ExporterStage}};
  
  \node (ExplorViz) [above=1.0cm of OpenTelemetryExporter] {ExplorViz};
  
  \node (Transformer) [above=0.2cm of MultipleConnectionTcpSource.west, xshift=-0.1cm, anchor=west, minimum width=16.1cm, minimum height=1.5cm] {\textbf{Kieker OpenTelemetry Transformer}\\[0.55cm]};
  
  \draw[->] (Probes)   -- (SingleSocketTcpWriter);
  \draw[->] (SingleSocketTcpWriter)   -- (MultipleConnectionTcpSource);
  \draw[->] (MultipleConnectionTcpSource)   -- (ExecutionRecordTransformation);
  \draw[->] (ExecutionRecordTransformation)   -- (TraceReconstruction);
  \draw[->] (TraceReconstruction)   -- (OpenTelemetryExporter);
  \draw[->] (OpenTelemetryExporter)   -- (ExplorViz);
  \end{tikzpicture}
  \end{center}
  \caption{Process of Exporting Kieker Traces with TeeTime Stages to ExplorViz}
  \label{fig:exportProcess}
\end{figure*}

Kieker's native records contain information about the program execution, for example points in the control flow and system information. Control flow points denote the beginning and the end of operations or thread joins, which are stored together with the respective timestamps. System information include the CPU and disk utilization. All of them are stored with respective metadata. The metadata of the control flow record allows the reconstruction of the call tree.

OpenTelemetry standardizes signals, which are typically traces, metrics and logs. A trace is a control flow of an application, which consists of potentially nested spans. Similar to Kieker, spans can denote operation execution but may also represent different units of work. A metric is a measurement of a service, which mirrors Kieker's system records. A log is a text with a timestamp, that might be structured. 

To introduce interoperability, we developed a transformation of Kieker's control flow records to OpenTelemetry traces. We used the pipe-and-filter framework TeeTime~\cite{ICSA2017TeeTime}. Kieker provides stages that consume TCP data and transform them into various Kieker formats. We added a stage that transforms Kieker records into an OpenTelemetry span. Since OpenTelemetry spans store their parent directly as a reference (via the \lstinline'setParent' method), the Kieker trace needs to be organized as a sorted list. Moreover, different types of Kieker records should potentially be the input of the transformation. Therefore, our \lstinline'OpenTelemetryExporterStage' accepts \lstinline'ExecutionTrace' instances as input, i.e., full reconstructed Kieker traces. These can be created from different types of Kieker records.
% Using the single \lstinline'Execution's from these traces, the individual spans are created.
Figure~\ref{fig:exportProcess} illustrates the role of the added stage in the overall process.
These spans are passed to an OpenTelemetry tracer. The OpenTelemetry SDK provides different export options. Currently, our \lstinline'OpenTelemetryExporterStage' supports export to gRPC (OpenTelemetrys default) and to Zipkin.
Our stage is used in an additionally added Kieker tool, the \lstinline'otel-transformer', which listens for TCP data. With minimal adaptions, the stage can be used in other contexts, e.g., for log replaying. 

\section{Case Study: Using Kieker Data for ExplorViz}

In this section, we describe the ExploreViz tool and our demo using it with TeaStore traces.

\subsection{ExplorViz}
ExplorViz is a web-based 3D software visualization tool which employs the city metaphor~\cite{hasselbring2020explorviz}.
It displays both the software structure and program behavior.
To collect runtime data, ExplorViz uses third party dynamic analysis tools which support the OpenTelemetry standard like NovaTec’s Java agent inspectIT Ocelot.\footnote{\url{https://www.inspectit.rocks}}
An OpenTelemetry Collector acts as a buffer and adapter between live tracing tools and the backend of ExplorViz.
The data is send via gRPC to the Collector and then send via Apache Kafka to and through the backend of ExplorViz.
Finally, the processed data is requested for ten second time periods and visualized for the user by the web-based frontend of ExplorViz.

\subsection{Demo}
The TeaStore~\cite{Eismann2018} is a reference microservice application that can be used for performance modeling and benchmarking. It consists of five microservices that interact with one another, all pre-instrumented with Kieker probes using AspectJ. To showcase our Kieker-to-OpenTelemetry export, we visualize tracing data from the TeaStore in ExplorViz.

To achieve this, we first utilize the existing Kieker probes and integrate a \lstinline'SingleSocketTcpWriter' to send the records to our transformer. Inside of the transformer, execution traces are reconstructed, and subsequently, the \lstinline'OpenTelemetryExporterStage' sends the records as spans to the OpenTelemetry Collector of ExplorViz.

\begin{figure}[htbp]
	\includegraphics[width=\textwidth/2]{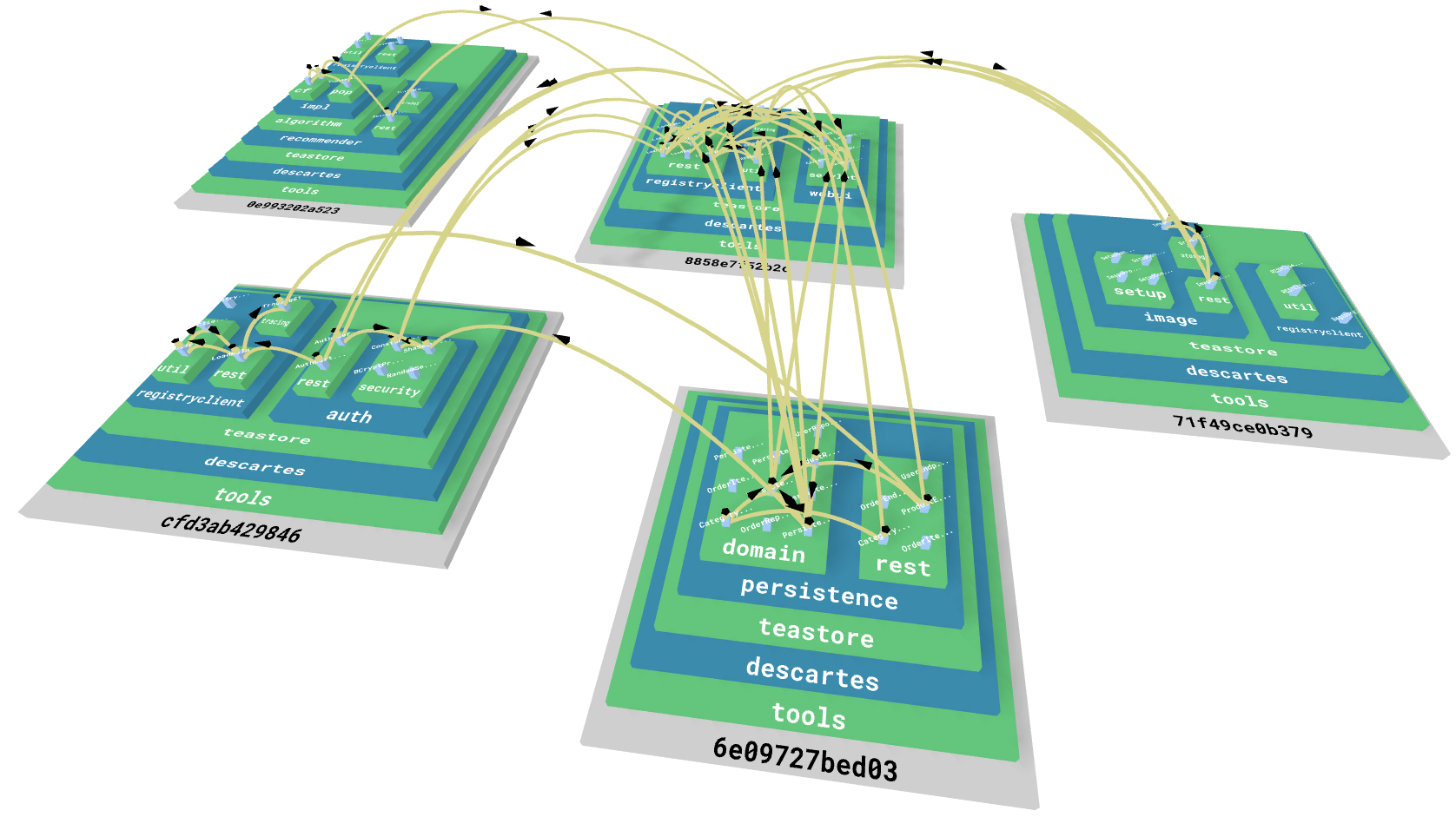}
        %TODO: Replace with screenshot from live visualization with readable service names
	\caption{TeaStore as visualized by ExplorViz}
	\label{fig:explorviz}
\end{figure}

The sent OpenTelemetry spans are aligned with the semantic conventions of OpenTelemetry and carry data such as application, package, class and method names as well as execution times of individual spans.
An example of ExplorViz's resulting visualization of the TeaStore is presented in Figure \ref{fig:explorviz}.
In this color scheme, gray foundations represent the five TeaStore services.
Nested green and blue boxes represent the packages as districts in the city metaphor, whereas classes are represented as light blue buildings.
Method calls which occurred within the selected time period are represented as yellow curved lines with arrows indicating the  direction of communication.

Popovers allow a user to display more detailed information about the visualized entities, for example, which method calls were recorded between two given classes.
The visualization can be adjusted to the user's needs through manifold settings and filter options.
Together with the ability to explore the visualization collaboratively, ExplorViz thereby aims to support program comprehension tasks by providing user's with a quick overview and details on demand about the visualized software system.

\section{Related Work}

Janes et al.~\cite{janes4175937open} examine interoperability of observability tools and find that all major observability tools, including AppDynamics, Datadog and Dynatrace, provide interoperability with the OpenTelemetry standard.

Furthermore, Okanovic et al.~\cite{okanovic2016towards} defined the OPEN.XTRACE format, which aimed to be a standard format for observability data like OpenTelemetry is today. They provided adapters for Dynatrace, inspectIT and Kieker. 
%The Kieker adapter was implemented as a Kieker plugin. 
They found that the interoperability of the tools was only partially given, since not all data were present in all observability tools. 
%For example, parameter values of method calls could at this point of time not be obtained by Kieker, but by Dynatrace. 
Their work can serve as a blueprint for future full OpenTelemetry interoperability. However, since OpenTelemetry became the de-facto industry standard, their implementations are not re-usable for interoperability.

Kunz et al.~\cite{kunz2017generic} research standardization of observability data by transforming them into performance models. They show that their Monitoring Data Transformation Platform (MTDT) is capable to generate performance models from observability data. 

\section{Summary and Outlook}

We provided a first step toward interoperability of Kieker and OpenTelemetry by providing an export of Kieker traces to OpenTelemetry. The core of our contribution is the \lstinline'OpenTelemetryExporterStage' that exports Kieker traces into OpenTelemetry via gRPC or Zipkin. We demonstrated our approach by exporting TeaStore records into ExplorViz.

To gain full interoperability of Kieker and OpenTelemetry, we are aiming for also importing OpenTelemetry traces into Kieker. Furthermore, another approach is to define the OpenTelemetry records by Kieker's Instrumentation Record Language (IRL) and to define appropriate writers that directly export OpenTelemetry records from Kieker's instrumentation. Finally, this future work can be concluded by comparing the overhead of writing Kieker records and transforming them, and the writing of OpenTelemetry records directly via Kieker.

\printbibliography

\end{document}